\documentclass[superscriptaddress,groupedaddress,nofootnoteinbib,11pt]{article}  
\pdfoutput=1

\usepackage{graphicx}  
\usepackage{dcolumn}   
\usepackage{bm}        
\usepackage{slashed}        
\usepackage{float}
\usepackage{jcappub}
\usepackage{color}

\usepackage{graphicx}
\usepackage{dcolumn}
\usepackage{bm}
\usepackage{amssymb}
\usepackage{amsmath}
\usepackage{sectsty}
\usepackage{colortbl}
\usepackage{latexsym}
\usepackage{float} 
\usepackage{ifthen}
\usepackage{enumerate}
\usepackage{url}
\usepackage[force]{feynmp-auto}
\usepackage{jcappub}
    \usepackage{picinpar}
    \usepackage{colortbl}
\usepackage{multirow}
	\usepackage{float}
	      \usepackage{setspace}
\usepackage{array}
\usepackage{bm}
\usepackage{amsopn}

\usepackage{booktabs}
\usepackage[table]{xcolor}

\def\be{\begin{equation}}
\def\ee{\end{equation}}
\def\ba{\begin{eqnarray}}
\def\ea{\end{eqnarray}}

\def\L*{{\cal L}_*}
\def\L{\mathcal{L}}
\def\({\left(}
\def\){\right)}

\def\<{\langle}
\def\>{\rangle}

\def\cs2{c_{s}^{2}}

\definecolor{verde}{rgb}{0,0.5,0}

\begin{document}

\title{Imprints of Massive Primordial Fields on Large-Scale Structure}

\author{Emanuela Dimastrogiovanni$^a$,}
\author{Matteo Fasiello$^b$,} \author{and Marc Kamionkowski$^c$}

\affiliation{$^a$Department of Physics and School of Earth and Space Exploration, Arizona State University, Tempe, AZ 85827-1404}
\affiliation{$^b$Stanford Institute for Theoretical Physics and Department of Physics, Stanford University, Stanford, CA 94306}
\affiliation{$^c$Department of Physics and Astronomy, 3400 N. Charles St., Johns Hopkins University, Baltimore, MD 21218 USA}

\date{\today}

\abstract{Attention has focussed recently on models of inflation
that involve a second or more fields with a mass near the inflationary
Hubble parameter $H$, as may occur in supersymmetric theories if
the supersymmetry-breaking scale is not far from $H$.
Quasi-single-field (QsF) inflation is a relatively simple family
of phenomenological models that serve as a proxy for theories
with additional fields with masses $m\sim H$. Since QsF inflation involves fields in addition to the inflaton, the consistency conditions between correlations that arise in single-clock inflation are not necessarily satisfied. As a result, correlation functions in the squeezed limit may be larger than in single-field inflation. Scalar non-Gaussianities mediated by the massive isocurvature field in QsF have been shown to be potentially observable. These are especially interesting since they would convey information about the mass of the isocurvature field. Here we consider non-Gaussian correlators involving tensor modes and their observational signatures. A physical correlation between a (long-wavelength) tensor mode and two scalar modes (\textsl{tss}), for instance, may give rise to local departures from statistical isotropy or, in other words, a non-trivial four-point function. The presence of the tensor mode may moreover be inferred geometrically from the shape dependence of the four-point function. We compute \textsl{tss} and \textsl{stt} (one soft curvature mode and two hard tensors) bispectra in QsF inflation, identifying the conditions necessary for these to ``violate" the consistency relations. We find that while consistency conditions are violated by \textsl{stt} correlations, they are preserved by the \textsl{tss} in the minimal QsF model. Our study of primordial correlators which include gravitons in seeking imprints of additional fields with masses $m\sim H$ during inflation can be seen as  complementary to the recent ``cosmological collider
physics'' proposal.}

\maketitle

\section{Introduction} 

Given that inflation with a single scalar degree of freedom is
already able to account for observations, one might wonder why
we should consider inflationary models with multiple fields.
One reason is that the flatness required of the inflaton potential
calls for a mechanism or symmetry to protect it from quantum
corrections. This is sometimes described as the \textit{eta
problem}.  

A supersymmetric UV completion is often invoked to
settle this issue.   If supersymmetry is not
broken at energies higher than the inflationary Hubble parameter
$H$, then the inflationary vacuum energy will be responsible for
breaking supersymmetry, since there is no supersymmetric theory
in de Sitter space. The theory becomes aware of the difference
between de Sitter and Minkowski spacetime at a scale
that probes effects from the curvature, which leads us directly
to $H$.  It follows that a natural mass range for fields
populating the supersymmetric multiplets will not stray too far
from $m \sim H$ \footnote{Again, the underlying assumption here
is that supersymmetry is being broken at $H$. If the breaking
scale were much higher, then the corresponding massive fields
may be integrated out, effectively reducing the dynamics to that
of a single degree of freedom.  Intriguingly, even in the
low-energy, single-field regime, one may find remnants of a past
multi-field dynamics (see e.g.,
Refs.~\cite{Tolley:2009fg,Achucarro:2012sm,Achucarro:2012yr,Burgess:2012dz}).}.

Recently, an interesting proposal was put forward
\cite{Chen:2009zp} (see also
Refs.~\cite{Craig:2014rta,Noumi:2012vr,Sefusatti:2012ye,McAllister:2012am}) under
the name of \textit{quasi single-field} (QsF) inflation.  
This paradigm is flexible enough to accommodate a number
of theories within its parameter space, yet constrained to
describe the dynamics of a light inflaton field with a number
of massive, $m \sim H$, scalar fields. 

This description sits in a somewhat convenient spot between
single-field inflation and fully multi-field models, where here
``multi-field models'' refers to theories with more
than one light field. Although it is possible that,  whatever
the mechanism responsible for protecting the mass of the
inflaton from quantum corrections, it might well do the same for
other fields, our perspective here will be that of limiting the
number of protected fields to the smallest necessary, namely
one. As we shall see, this choice comes with a number of
advantages, chief among which stands the more predictive nature
of quasi-single field inflation.

A field with a mass of order $H$ decays relatively quickly, so
much so that its most prominent (late-time) effects are those
mediated by its interactions\footnote{Indeed, in QsF inflation,
the linear equations of motion for the fields are typically
decoupled because the would-be mixing term in the Lagrangian is
assumed small and can be treated as a perturbation on top of
$\mathcal{L}_2$.} with the light inflaton field. It is the
``action-only-through-$\phi$'' (where $\phi$ is the inflaton)
that makes for a predictive theory and shields QsF inflation
from a host of highly model-dependent characterizations
extending to the post-inflationary evolution. This is in stark
contradistinction to what generally applies for fully
multi-field theories where the extra field(s) directly
contributes to late-time observables.

The non-Gaussianities in the scalar sector of QsF inflation have
been the subject of a thorough investigation in
Ref.~\cite{Chen:2009zp}. Here we will focus on 
correlations between scalar and tensor modes. A correlation between
a soft (long-wavelength) tensor mode and two short-wavelength curvature
fluctuations (henceforth \textsl{tss} correlation), for instance, would be instrumental in unveiling general features of inflationary models. Even though
it may be difficult to measure it directly, such a correlation
is bound to affect the two-point function of the cosmic
microwave background (CMB) fluctuations and galaxy distributions
$(i)$ by introducing a quadrupolar correction (if the
long-wavelength tensor mode is larger than the size of the
observable horizon); or $(ii)$ in the form of a local departure
from statistical isotropy for sub-Hubble long wavelength tensor
modes.  This local departure from statistical isotropy appears
in Fourier space as an off-diagonal correlation between
different Fourier modes of the density field, or equivalently, a
nontrivial four-point scalar correlation function.
These features, in their interpretation as remnants of
field dynamics during inflation, are often dubbed as
\textsl{fossils} \cite{Jeong:2012df,Dai:2013ikl,Dai:2013kra}. 

A \textsl{tss} correlation signal in standard single-field
slow-roll inflation (SFSR), and, more generally, in all
single-clock models,  is partially can¶celled by projection (late
times) effects, leaving an unobservably small (of order $\sim
k^2_{L}/k^2_{S}$) signal \cite{Pajer:2013ana,Dai:2013kra,Dai:2015rda}. This
is no longer the case for models that violate\footnote{It is important to note here what exactly is to be intended as``violation of ccs". It is not the non-linearly realized symmetry to be violated/broken. Rather, in all that follows, whenever a relation between a ``squeezed" $n+1$ and an n-point functions will require the knowledge of an additional non-trivial (i.e. not reducible to the original two ingredients) $n+1$-correlator (typically involving additional fields, see the ``extra" in Eq.(\ref{uuu})), we will say there is a ccs violation. } consistency
conditions as has been investigated in Ref.~\cite{Brahma:2013rua} for
inflation with non Bunch-Davies initial conditions,  in
Ref.~\cite{Dimastrogiovanni:2014ina} for inflation with a
non-attractor phase
\cite{Namjoo:2012aa,Chen:2013aj,Chen:2013eea} (see also
Ref.~\cite{Kinney:2005vj}), and for solid inflation
\cite{Endlich:2012pz,Endlich:2013dma,Endlich:2013jia,Akhshik:2014bla}.

In single-clock models of inflation, consistency conditions
(henceforth \textsl{ccs})
\cite{Maldacena:2002vr,Kehagias:2012pd,Creminelli:2012qr,Goldberger:2013rsa,Hinterbichler:2013dpa,Kehagias:2013xga,Berezhiani:2013ewa,Berezhiani:2014tda}
relate the $n$-point correlation functions to $(n-1)$-point
functions in the soft limit of one of the momenta.  They do not
necessarily apply, however, if there are multiple fields during inflation.
The squeezed limit of cosmological correlators is
therefore a powerful instrument to probe inflation.

In this paper we study \textsl{tss} and \textsl{stt} bispectra in QsF inflation in the squeezed limit for, respectively, the tensor and the scalar curvature mode. We show that the \textsl{stt} receives ccs-violating contributions, thus opening the way to new observables. On the other hand, observable signatures from \textsl{tss} correlations may not be generated within the minimal realization of QsF inflation as
introduced in Refs.~\cite{Chen:2009zp,Chen:2009we}. We show that, for a violation of ccs involving the \textsl{tss} correlation to occur, a non-zero two-point correlation between the soft tensor mode and the massive isocurvature mode is necessary. This is forbidden by the symmetries of the theory in the minimal QsF scenario. However, such violation may arise in non-minimal realizations with broken statistical isotropy or in the presence of additional vector or tensor modes.

This work is organized as follows: in Section \ref{review} we
review the main features of the model and its predictions for
the scalar sector; in Section \ref{sec.three} we provide a general discussion on signatures from a tensor-scalar-scalar correlator; in Section \ref{sec:four} we present our findings on the three-point correlation functions involving gravitons; in Section \ref{conclusions} we offer comments and conclusions. The
details of some of our derivations are provided in Appendix~\ref{diagramf}.

\section{The model: review of background and perturbation analysis}
\label{review}

QsF inflation comprises a class of multi-field models
characterized by the presence of a light field (the curvature
mode) driving inflation, and one or more fields
(isocurvature modes) with masses of order of the Hubble scale
$H$. Curvature and isocurvature modes couple in the presence of
a turning trajectory in field space \cite{Gordon:2000hv}. In
Ref.~\cite{Chen:2009zp,Chen:2009we}\footnote{See also
Refs.~\cite{Baumann:2011su,Chen:2012ge,Pi:2012gf,Assassi:2013gxa} for
further studies on this model and Ref.~\cite{Baumann:2011nk} for the
connection between supersymmetry and Hubble-mass degrees of
freedom (a recent model with e.g. $m_{\rm axion}\sim H$ is found in \cite{Chung:2015pga}).} a very simple quasi-single-field model was analyzed,
with one massive field ($\sigma$) only, and a minimal coupling
to gravity. Using polar coordinates in field space, the
Lagrangian has the form,
\begin{eqnarray}\label{21}
     \mathcal{S}_{m}=\int d^{4}x
     \sqrt{-g}\left[-\frac{1}{2}
     \left(\tilde{R}+\sigma\right)^{2}g^{\mu\nu}(\partial_{\mu}\theta)
     (\partial_{\nu}\theta)-\frac{1}{2}g^{\mu\nu}
     (\partial_{\mu}\sigma) ( \partial_{\nu}\sigma)
     -V_{sr}(\theta)-V(\sigma)\right],
\end{eqnarray}
where $\tilde{R}\theta$ corresponds to the adiabatic direction,
parallel to the inflaton field trajectory, and $\sigma$ to the
perpendicular (isocurvature) direction. For simplicity, the
trajectory in field space is assumed to have constant radius and
to be characterized by a constant angular velocity
(\textsl{constant turn}). Another simplifying assumption
concerns the magnitude of interactions between $\sigma$ and
$\theta$ at the quadratic level:
\begin{eqnarray}
     \mathcal{L}_{C_2}=2 a^{3} R \, \dot{\theta}_{0}
     \delta\sigma\delta\dot{\theta},
\end{eqnarray}
where we have replaced $R=\tilde{R}+ \sigma_0$, with $\sigma_0$
being the value of $\sigma$ at the minimum of its effective
potential.

\begin{figure}
\begin{center}
\includegraphics[scale=0.35]{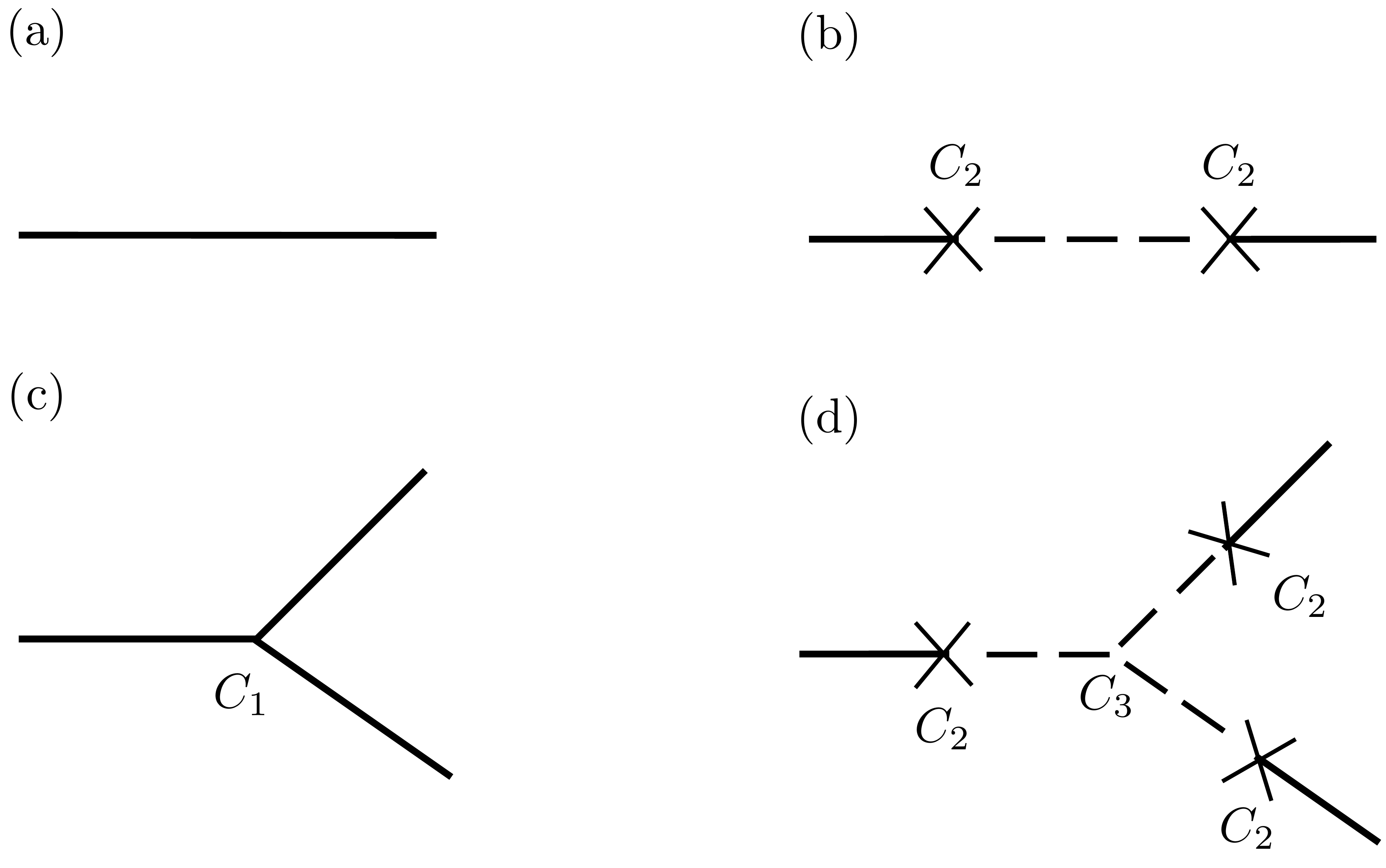}
\caption{ (a) Standard power spectrum of $\zeta$. (b) Correction to the power spectrum of $\zeta$ from interactions with $\sigma$. (c) Bispectrum of $\zeta$ from self-interactions. (d) Bispectrum of $\zeta$ from interactions with $\sigma$. Dashed line are associated with $\sigma$, continuous ones with $\zeta$.  }
\label{fig1}
\end{center}
\end{figure}

For small values of the coupling, $|\dot{\theta}_{0}/H|\ll 1$,
one can treat $\mathcal{L}_{C_2}$ as a perturbation on top of
the second-order Lagrangian for the two decoupled free
fields. This allows us to separately solve for the linear
fluctuations and then study the mixing between adiabatic and
isocurvature modes using perturbation theory. The mixing has
been ``postponed'' up until when interactions are considered. The
free-field mode functions have the form: 
\begin{eqnarray}\label{oneeq}
&&\delta\theta_{k}=\frac{H}{R\sqrt{2 k^{3}}}(1+i k\tau)e^{-i k\tau}\,,\\\label{twoeq}
&&\delta\sigma_{k}=-i e^{i(\nu+1/2)\pi/2}\frac{\sqrt{\pi}}{2}H(-\tau)^{3/2} H_{\nu}^{(1)}(-k\tau)\quad\quad \text{for}\,\,\,m \leq\frac{3\, H}{2}\,,
\end{eqnarray}
where $\nu\equiv\sqrt{9/4-m^{2}/H^{2}}$, and $m^{2}\equiv
V^{''}+7\,\dot{\theta}_{0}^{2}$ is the mass of the isocurvature
fluctuation\footnote{The $m>(3/2)\, H$ case entails an oscillatory
behavior at late times which can suppress the contribution from
$\delta\sigma$-mediated correlators. For this reason the authors
of \cite{Chen:2009zp,Chen:2009we} focus on the $m\leq (3/2)\,H$
case. Note however that the lower part of this mass range is
still far from the integrating out regime and supports a very
interesting dynamics \cite{Green:2013rd,
Arkani-Hamed:2015bza}.}. As expected, the larger the mass the
faster the mode function of the isocurvature field asymptotes to
zero after horizon crossing,
\begin{eqnarray}\label{asb}
\delta\sigma_{k}(\tau) \sim \left(-\tau\right)^{3/2-\nu}\,,\quad\quad (k\tau\,\rightarrow\,0)\,.
\end{eqnarray}
At tree level, the power spectrum of curvature fluctuations
receives a scale-invariant contribution from interactions of the
type in $\mathcal{L}_{C_2}$ (as represented in diagram (b)
of Fig.\ref{fig1}) that add up to the standard power spectrum
(diagram (a) of the same figure):
\begin{eqnarray}\label{26}
\mathcal{P}_{\zeta}=\frac{H^{4}}{4\pi^{2}R^{2}\dot{\theta}_{0}^{2}}\left[1+8\,\mathcal{C}(\nu)\frac{\dot{\theta}_{0}^{2}}{H^{2}}\right]\,,\quad\quad\quad n_{s}-1=-2\epsilon-\eta+8\,\eta\,\mathcal{C}(\nu)\frac{\dot{\theta}_{0}^{2}}{H^{2}}\,,
\end{eqnarray}
where $\mathcal{C}$ is a function of the mass of the
isocurvature field (see Refs.~\cite{Chen:2009zp,Chen:2009we} for its
numerical evaluation) and $\zeta\simeq
-H\,\delta\theta/\dot{\theta}_{0}$.

The field $\sigma$ is not in slow-roll (it sits in the minimum
of the effective potential) and, as a consequence, the magnitude
of $V^{'''}$ can be much larger than the potential in
conventional SFSR scenarios or in the case of multiple light
fields models.  As a result, contributions to the scalar
bispectrum arise from diagrams that are mediated by
$\delta\sigma$ (as in diagram (d) of Fig.~\ref{fig1}) that can
be large compared to the standard one-vertex diagram (diagram
(c) of Fig.~\ref{fig1}). One can see that the magnitude of
diagram (d) would be suppressed if $\sigma$ itself was in
slow-roll:
\begin{eqnarray}\label{itsr}
\mathcal{L}_{C_3}=-\frac{ a^{3}}{6}\,V^{'''}\delta\sigma^{3}.
\end{eqnarray} 
In the squeezed limit, $k_{3}\ll k_{1}\simeq k_{2}$, and for
$\nu\ne 0$, the contribution to the scalar bispectrum from
diagram (d) is found \cite{Chen:2009zp,Chen:2009we} to have the
following momentum dependence on the mass of the isocurvature
fluctuations
\begin{eqnarray}\label{mdep}
\langle \zeta_{\vec{k}_{1}}\zeta_{\vec{k}_{2}}\zeta_{\vec{k}_{3}}\rangle \sim \frac{1}{k_{1}^{3}k_{3}^{3}}\left(\frac{k_{3}}{k_{1}}\right)^{\frac{3}{2}-\nu}\,.
\end{eqnarray}
The momentum dependence in Eq.~(\ref{mdep}) is easily understood
from the super-horizon decay of the isocurvature fluctuations,
as given in Eq.~(\ref{twoeq}): in models with standard
Bunch-Davies initial conditions, as is the case for QsF
inflation in Ref.~\cite{Chen:2009zp,Chen:2009we}, a correlation
between long and short wavelengths can only be generated once
the short-wavelength modes approach horizon scales; by that
time, the amplitude of the long-wavelength mode will have
decayed by
$(\tau_{1}/\tau_{3})^{3/2-\nu}=(k_{3}/k_{1})^{3/2-\nu}$.

The shape function in Eq.~(\ref{mdep}) interpolates between
local (to which it reduces for $\nu\rightarrow 3/2$) and
 equilateral (approximately approached for $\nu\rightarrow 0$). The interpolation between
local and equilateral shapes (see Ref.~\cite{Chen:2009we} for more
details) can be easily understood: the more massive
$\delta\sigma$ is, the faster it decays on super-horizon scales
and therefore, the larger the contribution to non-Gaussianity
generated around horizon-crossing scales (equilateral type). On
the other hand, for lighter masses of $\delta\sigma$, the
super-horizon isocurvature fluctuations survive longer and can
contribute to correlations between long and short modes.  \\

Let us briefly comment on the prospects for detection of the scalar bispectrum in QsF inflation. If $V^{'''}/H\lesssim \mathcal{O}(1)$, the leading contribution arises from the interaction in Eq.~(\ref{itsr}). The amplitude reads \cite{Chen:2009zp}:
\begin{equation}\label{add11}
f_{NL}\sim \frac{1}{\sqrt{\mathcal{P}_{\zeta}}}\frac{V^{'''}}{H}\left(\frac{\dot{\theta}}{H}\right)^{3}\,.
\end{equation}
While we know that $f_{NL}$ in Eq.~(\ref{add11}) could be of order 10 (for the largest possible $V^{'''}$ values, the constraint being $V^{'''}< H$), it is just as true that it could be much smaller (and yet dominated by the $V^{'''}$ contribution). It follows that such a signal may or may not be accessible for, e.g., upcoming LSS surveys. 

\section{Tensor-scalar-scalar correlation and observational signatures}
\label{sec.three}

A correlation between one long-wavelength tensor and two
short-wavelength scalar modes manifests itself in the
distribution of primordial density fluctuations as an off-diagonal
correlation between different Fourier modes of the density
field. This is true for any primordial field that correlates
with curvature fluctuations in the squeezed limit
during inflation \cite{Jeong:2012df,Dai:2013ikl,Dai:2013kra},
but the nature of the correlation depends on the spin of the
field.  In the presence of a long-wavelength mode of the new
field, the correlation between two different Fourier modes becomes,
\begin{eqnarray}\label{a}
     \langle \delta(\vec{k}_{1})\delta(\vec{k}_{2})
     \rangle_{h^{p}(\vec{K})} =
     (2\pi)^{3}\left[\delta^{(3)}_{\vec{k}_{1}\vec{k}_{2}}
     P(k_{1}) + \delta^{(3)}_{\vec{k}_{1}\vec{k}_{2}\vec{K}}
     f_{p}(\vec{k}_{1},\vec{k}_{2}) h_{p}^{*}(\vec{K})
     \epsilon_{ij}^{p}(\hat{K})\hat{k}_{1i}\hat{k}_{2j}\right],  
\end{eqnarray}
where $h^{p}$ is the Fourier transform of the field, $p$ its
polarization, $\delta^{(3)}_{\vec{k}_{1}...\vec{k}_{n}}$
indicates the delta function of argument
$\vec{k}_{1}+...+\vec{k}_{n}$. The function $f_{p}$ is related
to the three-point correlation of the field with the curvature
fluctuations by 
\begin{eqnarray}
     B_{p}(\vec{K},\vec{k}_{1},\vec{k}_{2}) =
     P_{p}(K)f_{p}(\vec{k}_{1},\vec{k}_{2})\epsilon_{ij}^{p}(\hat{K})
     \hat{k}_{1i}\hat{k}_{2j }=
     \mathcal{B}_{p}(K,k_{1},k_{2})
     \epsilon_{ij}^{p}(\hat{K})\hat{k}_{1i}\hat{k}_{2j},
\end{eqnarray}
where $P_{p}$ is the power spectrum. Global statistical isotropy
requires that $f$ is only a function of the norm of
$\vec{k}_{1}$ and $\vec{k}_{2}$ and of the angle between the two
vectors.

Whenever the long-wavelength mode is super-Hubble, $k_{1}$ and
$k_{2}$ are indistinguishable from one another.  In this case, the
two contributions on the right-hand side of Eq.~(\ref{a}) can be
condensed into a single diagonal term that corresponds to the
sum of the regular scalar power spectrum plus an anisotropic
correction,
\begin{eqnarray}\label{a1}
     P(\vec{k})\simeq
     P(k)\left[1+\alpha\,\gamma_{ij}
     \hat{k}_{i}\hat{k}_{j}\right],
\end{eqnarray}
where $\gamma_{ij}$ is the tensor perturbation in our observable
volume.  Here $\alpha$ is related to the amplitude of the
\textsl{tss} correlation, $\alpha\sim
\mathcal{B}(k_{L},k_{S},k_{S})/P(k_{S})P_{\gamma}(k_{L})$.   Eq.~(\ref{a1}) shows that
the expectation value $\left\langle \left|\delta(\vec
k)\right|^2 \right \rangle$
about the direction ${\hat K}$ of the long-wavelength mode can
be isotropic (if the extra field is a scalar) or anisotropic if
the extra field is a vector or tensor field
\cite{Jeong:2012df,Dai:2013ikl}. In this work, the field $h^{p}$
is the tensor mode from the metric ($\gamma^{p}$).

Null searches for the quadrupolar anisotropies described by
Eq.~(\ref{a1}), when the new field is a tensor field, both in
the CMB
\cite{Pullen:2007tu,Hanson:2009gu,Groeneboom:2008fz,Bennett:2010jb,Ade:2013nlj,Kim:2013gka}
and LSS \cite{Ando:2008zza,Pullen:2010zy}, have resulted in a lower
bound $\lesssim \mathcal{O}(0.01)$ on the magnitude of the quadrupole.

\begin{figure}
\begin{center}
\includegraphics[scale=0.40]{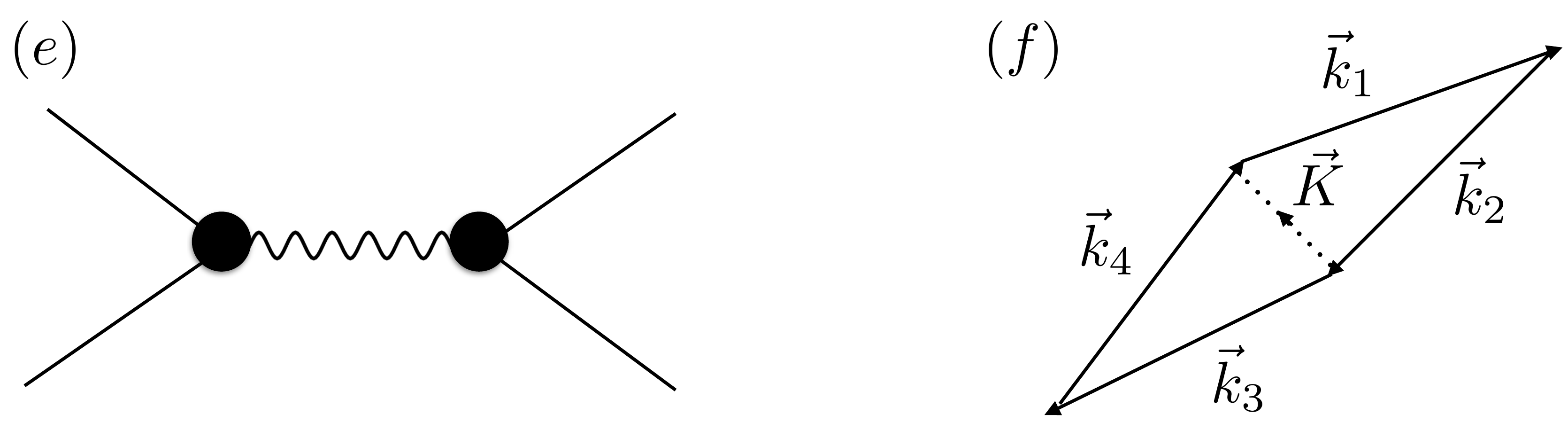}
\caption{(e)  Representation of a graviton-exchange diagram in the four-point function (which we mimic in the soft $K\ll k_i$ limit). Note the black shaded area stands for a generic type of interaction. (f) A pictorial representation of the momenta configuration of the non-trivial four-point function we are effectively probing in Eq.~(\ref{vara}).}
\label{fig_new_1}
\end{center}
\end{figure}

If the long-wavelength mode is sub-Hubble, any set of two scalar
modes of wavenumbers $\vec{k}_{1}$ and $\vec{k}_{2}$ with $\vec
k_1+\vec k_2+\vec{K}=0$ can be used to estimate the amplitude of
the tensor mode, providing a powerful probe that may well
complement other existing ones aimed at the search for
primordial gravitational waves (including
measurements of the B-mode CMB polarization
\cite{Seljak:1996gy,Kamionkowski:1996zd,Ade:2014xna,Ade:2014gua,Ade:2014afa,Ade:2015tva},
gravitational lensing effects in the CMB
\cite{Dodelson:2003bv,Dai:2012bc,Cooray:2005hm,Pi:2012gf}, LSS
\cite{Cooray:2005hm,Li:2006si,Dodelson:2010qu,Book:2011na,Schmidt:2012nw}
and 21cm cosmology \cite{Pen:2003yv,Masui:2010cz,Book:2011dz},
and direct gravitational-wave searches
\cite{Liddle:1993zj,BarKana:1994bu,Turner:1996ck,Smith:2005mm,Smith:2008pf,Chongchitnan:2006pe,Kuroyanagi:2014qaa,Jinno:2014qka}). For
a stochastic GW background, the minimum-variance estimator for
the tensor amplitude is \cite{Jeong:2012df},
\begin{eqnarray}\label{vara}
\hat{A}_{\gamma}=\sigma_{\gamma}^{2}\sum_{\vec{K},p}\frac{\left(P_{\gamma}^{f}(K)\right)^{2}}{2\left(P_{p}^{n}(K)\right)^{2}}\left(\frac{|\hat{\gamma}_{p}(\vec{K})|^{2}}{V}-P_{p}^{n}(K)\right)\, ,
\end{eqnarray}
where $P^{f}_{p}\equiv P_{\gamma}(K)/A_{\gamma}$ is a fiducial
power spectrum for the tensor modes. The quantity
$\hat{\gamma}_{p}(\vec{K})$ is the optimal estimator for the
amplitude of a single Fourier mode,
\begin{eqnarray}
\hat{\gamma}_{p}(\vec{K})\equiv P_{p}^{n}(K)\sum_{\vec{k}}\frac{B_{p}(\vec{K},\vec{k},\vec{K}-\vec{k})/P_{\gamma}(K)}{2VP^{\rm tot}(k)P^{\rm tot}(|\vec{K}-\vec{k}|)}\delta(\vec{k})\delta(\vec{K}-\vec{k})\, ,
\label{gamma3}
\end{eqnarray}
and $P_{p}^{n}(K)$ its variance
\begin{eqnarray}\label{riprendi}
P_{p}^{n}(K)\equiv \left[\sum_{\vec{k}}\frac{|B_{p}(\vec{K},\vec{k},\vec{K}-\vec{k})/P_{\gamma}(K)|^{2}}{2VP^{\rm tot}(k)P^{\rm tot}(|\vec{K}-\vec{k}|)}\right]^{-1}.
\end{eqnarray}
In the above formulas $V\equiv(2\pi/k_{\rm min})^{3}$ stands for
the volume of the survey and $P^{\rm tot}$ is the total scalar
power spectrum, including signal and noise. The variance of the
estimator in Eq.~(\ref{vara})
\begin{eqnarray}
\sigma^{-2}_{\gamma}\equiv\frac{1}{2}\sum_{\vec{K},p}\left(P^{f}_{p}(K)^{2}/P_{p}^{n}(K)^{2}\right)\, ,
\end{eqnarray} 
may be used to estimate the smallest GW amplitude that can be
detected for a given survey size.

As already mentioned, the actual observed quantities are found
after an additional step. Specifically, one needs to subtract
from the primordial \textsl{tss} correlation a term accounting
for late-time effects
\cite{Dai:2013kra,Pajer:2013ana,Dai:2015rda},
\begin{eqnarray}\label{obb}
     \mathcal{B}_{obs}(k_{L},k_{S},k_{S})=\mathcal{B}(k_{L},k_{S},k_{S})
     - \mathcal{B}_{\rm cc}(k_{L},k_{S},k_{S}),
\end{eqnarray}
where the \textsl{tss} correlator has the generic form,
\begin{eqnarray}
\langle \gamma_{\vec{k}_{1}}^{p}  \zeta_{\vec{k}_{2}}\zeta_{\vec{k}_{3}}\rangle=(2\pi)^{3}\delta^{(3)}_{\vec{k}_{1}\vec{k}_{2}\vec{k}_{3}}\epsilon^{p}_{ij}(\hat{k}_{1})\hat{k}_{2i}\hat{k}_{3j}\mathcal{B}(k_{1},k_{2},k_{3}),
\end{eqnarray}
and we set $k_{L}\equiv k_{1}\ll k_{2},k_{3}\simeq k_{S}$. The
second term on the right-hand side of Eq.~(\ref{obb}) corresponds
precisely to the expression, to leading order in powers of
$(k_{L}/k_{S})$, for the squeezed limit \textsl{tss} correlator
as dictated by consistency conditions \cite{Maldacena:2002vr}:
\ba\label{310}
\mathcal{B}_{\rm cc}(k_{L},k_{S},k_{S})\equiv -\frac{1}{2}P_{\gamma}(k_{L})P_{\zeta}(k_{s})\frac{\partial \ln P_{\zeta}(k_{S})}{\partial \ln k_{S}}\,.
\ea
In models where the consistency conditions are satisfied, the
leading-order part of the primordial signal would then be, for
the most part, cancelled by projection effects, leaving an
unobservably small $\mathcal{O}(k^2_{L}/k^2_{S})$ signal. As a
result, observations (either direct or indirect) of a
tensor-scalar-scalar correlation would rule out single-clock
models of inflation.\\

Eq.~(\ref{vara}) and (\ref{gamma3}) reveal that we are effectively probing a specialized four-point function in the limit that approaches the so-called counter-collinear configuration, see Fig.~\ref{fig_new_1}(e).

\begin{figure}
\begin{center}
\includegraphics[scale=0.35]{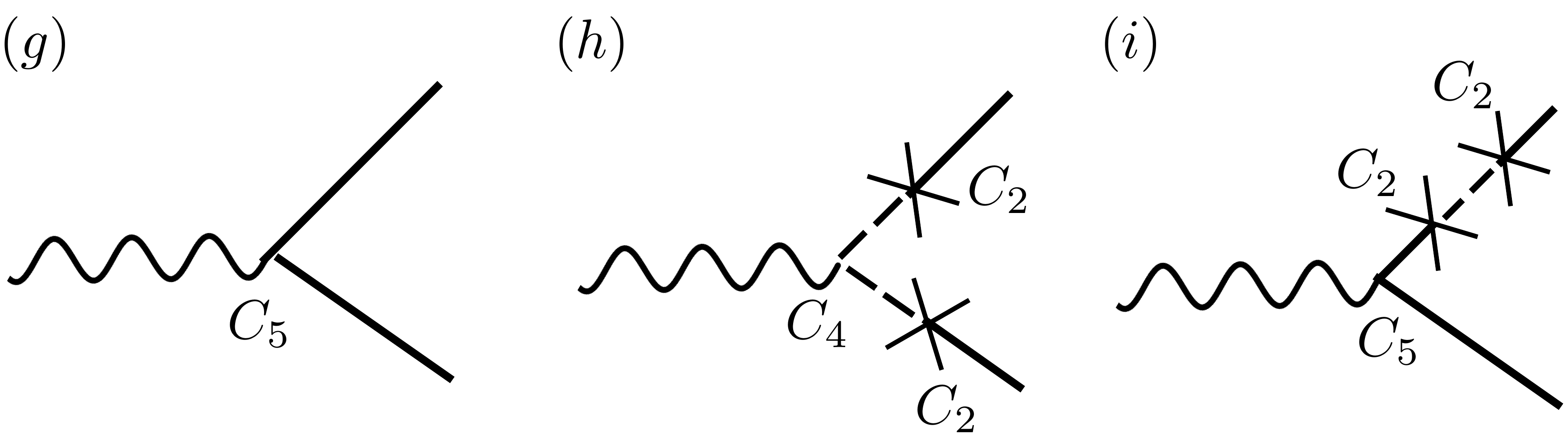}
\caption{(g) Tensor-scalar-scalar correlation from usual coupling between $\gamma$ and $\zeta$. (h,i) Leading order tensor-scalar-scalar correlator mediated by $\sigma$. Dashed line are associated with $\sigma$, wiggly lines with $\gamma$ and solid lines with $\zeta$.}
\label{fig_new_2}
\end{center}
\end{figure}

\section{Correlators with tensors in QsF inflation}
\label{sec:four}

\begin{figure}
\centering
   \includegraphics[width=.6\linewidth]{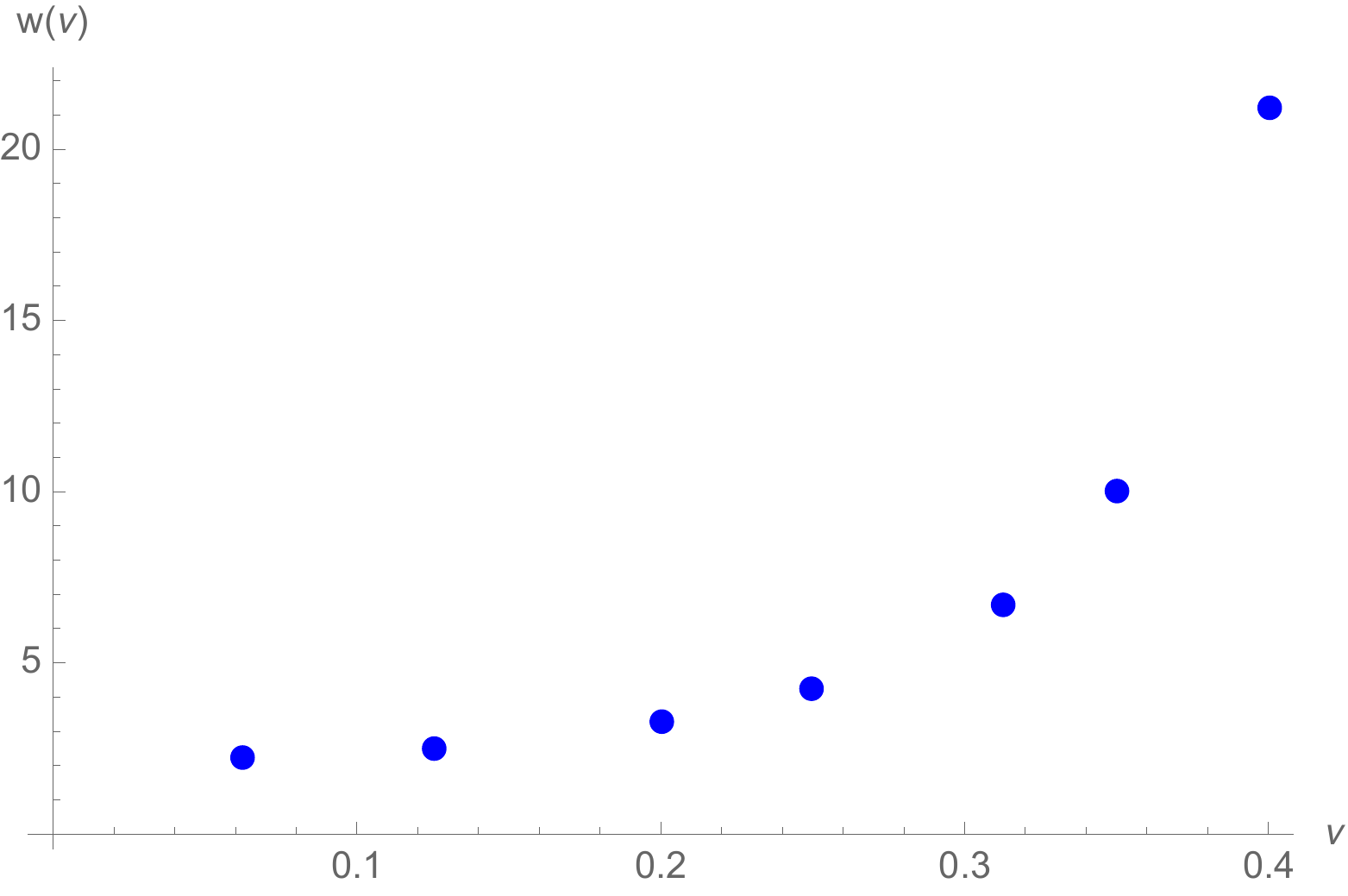}
\caption{Plot of numerical values of the coefficient $w(\nu)$
introduced in Eq.~(\ref{res11}) for $\nu$ ranging from $0.06$ up
to $0.4$.}
\label{gg}
\end{figure}

Similarly to the scalar bispectrum, the tensor-scalar-scalar three-point function may be generated 
by direct interaction of tensor modes with scalar curvature (as in diagram (g) of Fig.~\ref{fig_new_2}) 
\begin{eqnarray}
\mathcal{L}_{C_5}=\frac{a^{}}{2}\frac{R^2 \dot{\theta}^{2}}{H^{2}}\,\gamma_{ij}\partial_{i}\zeta\partial_{j}\zeta\,,
\end{eqnarray}
or as the result of interactions between the external fields and the isocurvature mode (as in diagram (h) and (i) of Fig.~\ref{fig_new_2}). 
Diagram (h) arises from the interactions between $\delta\sigma$
and the graviton, as well as from the interactions between
$\delta\sigma$ and $\zeta$:
\begin{eqnarray}
\mathcal{L}_{C_2}= -2\,a^{3} \frac{R\dot{\theta}_{0}^{2}}{H}\delta\sigma\,\dot{\zeta}\,,\quad\quad\quad\quad
\mathcal{L}_{C_4}=\frac{a}{2}\gamma_{ij}\partial_{i}\delta\sigma\partial_{j}\delta\sigma\,.
\end{eqnarray}
The diagram can be computed using the in-in formalism
\cite{Schwinger:1960qe}. We refer the reader to
Appendix~\ref{diagramf} for the details of the derivation and
report here the final result in the squeezed limit
\begin{eqnarray}\label{res11}
\mathcal{B}_{(h)}(k_{L},k_{S},k_{S})\simeq -\frac{\pi^{2}}{2}\,w(\nu)\frac{\dot{\theta}_{0}^{2}}{H^{2}}\left(\frac{H^{2}}{M_{P}^{2}k_{L}^{3}}\right)\left(\frac{H^{4}}{2\dot{\theta}_{0}^{2}R^{2}k_{S}^{3}}\right)=-\frac{\pi^{2}}{2}\,w(\nu)\frac{\dot{\theta}_{0}^{2}}{H^{2}}P_{\gamma}(k_{L})P_{\zeta}^{(0)}(k_{S})\,,\nonumber\\
\end{eqnarray}
 Here,
$P_{\gamma}(k)\equiv(H^{2})/(M_{P}^{2}k^{3})$ is the tensor
power spectrum, and $P_{\zeta}^{(0)}(k)\equiv(H^{4})(2
R^{2}\dot{\theta}_{0}^{2}k^{3})$ the leading order part of the
scalar power spectrum. In order to maintain analytical control, we have calculated the \textsl{tss} correlator directly in the squeezed limit.  One could schematically think of the complete \textsl{tss} correlator as, for example, the sum of two contributions, one that dominates in the squeezed limit and the other that dominates, e.g., in the equilateral limit. However, our variance in Eq.~(\ref{vara}) accounts for a hierarchy between momenta associated with tensors and scalar modes, it is built from an observable evaluated in the squeezed limit, therefore one would expect that it would not be severely affected by contributions from non-squeezed momenta configurations. 
The function $w(\nu)$ is shown in Fig.~\ref{gg} for a
range of $\nu$ values. We defined $k_{L}$ and $k_{S}$ as,
respectively, the (tensor) long- and (scalar) short-wavelength
modes.

{The \textsl{tss} correlation here is of the local
type: no suppression as that appearing in Eq.~(\ref{mdep}) is
now present  because $\delta\sigma$ does not effectively carry
the soft momentum. One can see this at the level of the Feynman diagrams for the IN-IN formalism employed in our calculations. Whilst in Fig.~\ref{fig1}(d) one isocurvature fluctuation $\delta\sigma$ is necessarily contracted with an external scalar fluctuation, forcing its wavenumber to be soft due to momentum conservation, this is not the case of Fig.~\ref{fig_new_2}(h),  where the external tensor attaches directly to a three-point vertex. Correlations in the squeezed
limit are generated as soon as the isocurvature modes have
reached horizon size, before they undergo any damping.}

Notice that the r.h.s. of Eq.~(\ref{310}) can be expanded to next-to-leading order in slow-roll: in this case it will include contributions proportional to $(\dot{\theta}/H)^{2}$ (see e.g. the correction from the isocurvature mode to the curvature power spectrum in Eq.~(\ref{26})). This observation, along with our findings in Eq.~(\ref{res11}), are suggestive of the fact that consistency conditions may be preserved\footnote{We are grateful to L. Bordin, P. Creminelli, J. Nore\~na for illuminating discussions on this point.} for the \textsl{tss} correlator and they should be carefully checked.

This can be done by computing diagram (i) of Fig.~\ref{fig_new_2}. It is easy to verify that, modulo an overall numerical coefficient (in the form of some function, $g(\nu)$, of the isocurvature field mass), diagrams (h) and (i) have the same parametric and momentum dependence. One may then follow the steps outlined in Appendix~\ref{diagramf} to check the consistency conditions with a straightforward but lengthy calculation. Given the approximations and the final numerical integration involved in our squeezed-limit computation, this may not be the most convenient route towards verifying an exact cancellation. We opt instead for a more general procedure that allows us to also identify what is to be expected of a given model for the consistency conditions to be violated. We outline this in the following subsection.

\subsection{Consistency conditions for inflation with extra fields}

{We will be following the notation and logic of \cite{Creminelli:2012ed,Assassi:2012zq,Hinterbichler:2013dpa} to show that the general form of the tensor-scalar-scalar correlator, to leading order in the squeezed limit, is as follows
\begin{equation}\label{uuu}
\lim_{\vec{k}_{1}\rightarrow 0}\frac{\big\langle \gamma^{s}_{\vec{k}_{1}}\zeta_{\vec{k}_{2}}\zeta_{\vec{k}_{3}}\big\rangle^{'}_{c}}{P_{\gamma}(k_{1})}=-\frac{1}{2} \epsilon_{ij}^{s}(\hat{k}_{1})\sum_{a=2}^{3}k_{a}^{i}\frac{\partial}{\partial k_{a}^{j}}\big\langle\zeta_{\vec{k}_{2}}\zeta_{\vec{k}_{3}}\big\rangle^{'}_{c}+[\text{extra}]\,,
\end{equation}
where a contribution proportional to the first one on the right-hand side is the only one found when, in our nomenclature, the model preserves consistency conditions \footnote{Here  $[extra]$ stands for everything that makes up the relation dictated by the symmetry. It is important to stress that, in general, only a subset of  $[extra]$ violates the ccs. The culprits are in particular quantities, such as some of those contributing to the last term in Eq.~(\ref{rrrrr}), which cannot be trivially reduced to the usual (l.h.s. and r.h.s.) two terms in Eq. (\ref{uuu}) above. }. We will briefly discuss the origin and form of the extra contribution that appears when these are violated. In Eq.~(\ref{uuu}), a prime stands for an (omitted) momentum-conserving delta function and a subscript ``c'' indicates connected diagrams. We will drop the connected diagram subscript from here on.\\
First off, one needs to identify the (non-linearly realized) symmetry that generates the soft-tensor consistency conditions. This corresponds to an anisotropic spatial rescaling of coordinates {
\ba
x^{i}\rightarrow x^{i}+S_{ij}\,x^{j}\; ,
\ea
}
where $S_{ij}$ is a symmetric, transverse and traceless tensor. Under this coordinate transformation the curvature fluctuation shifts linearly ($\delta_{\omega}\zeta\sim S_{ij}\,x^{j}\partial_{i}\zeta$), whereas crucially the tensor modes transformation has a non-linear contribution ($\delta_{\omega}\gamma_{ij}\sim S_{ij}$). Introducing the Noether charge, $Q_{\omega}$, associated with the symmetry transformation, one finds:
\begin{equation} \label{rrtt}
\big\langle \delta_{\omega}\gamma_{ij}(\vec{q})\big\rangle= i\, \big\langle \left[Q_{\omega},\gamma_{ij}(\vec{q})\right]\big\rangle\sim (2\pi)^{3}\delta^{(3)}(\vec{q}) S_{ij}\,.
\end{equation}
Consider now the effect of the transformation on the operator $\zeta_{\vec{k}_{1}}\,\zeta_{\vec{k}_{2}}$ while, at the same time, introducing a complete set of states $n_{\vec{k}}$ : 
\begin{equation}\label{1pa}
\big\langle \left[Q_{\omega},\zeta_{\vec{k}_{2}}\zeta_{\vec{k}_{3}}\right]\big\rangle=\sum_{n,\vec{k}}\big\langle Q_{\omega}| n_{\vec{k}}\big\rangle\big\langle n_{\vec{k}} | \zeta_{\vec{k}_{2}}\zeta_{\vec{k}_{3}}\big\rangle-\sum_{n,\vec{k}}\big\langle \zeta_{\vec{k}_{2}}\zeta_{\vec{k}_{3}} | n_{\vec{k}}\big\rangle\big\langle n_{\vec{k}} | Q_{\omega}\big\rangle\,.
\end{equation}
We define one-particle states generated by $\zeta$, $\gamma$ and by the orthogonal projection of the isocurvature mode $\tilde\sigma$ as: 
\begin{equation}
 |1_{\zeta_{\vec{k}}}\rangle \equiv P_{\zeta}^{-1/2}(k) |\zeta_{\vec{k}}\rangle    \,, \quad\quad | 1_{\gamma^{p}_{ij,\vec{k}}}\rangle \equiv P_{\gamma}^{-1/2}(k) | \gamma^{p}_{ij,\vec{k}} \rangle  \,, \quad\quad  | 1_{\tilde{\sigma}_{\vec{k}}}\rangle\equiv P_{\tilde\sigma}^{-1/2}(k)| \tilde{\sigma}_{\vec{k}}\rangle\,,
\end{equation}
where 
\ba
\tilde\sigma_{\vec{k}}\equiv \sigma_{\vec{k}}- P_{\zeta}^{-1}(k)\langle \zeta_{\vec{k}}\, \sigma_{\vec{k}} \rangle^{'}\zeta_{\vec{k}}- P_{\gamma}^{-1}(k)\langle \gamma^{p}_{ij,\vec{k}} \,\sigma_{\vec{k}} \rangle^{'}\gamma^{p}_{ij,\vec{k}}\; ,
\ea

\noindent {and we have used the fact that $\langle \zeta \gamma \rangle = 0 $ by rotational invariance.} Using this in Eq.~(\ref{1pa}) we obtain: 
\begin{eqnarray}\label{reee}
\Big\langle \left[Q_{\omega},\zeta_{\vec{k}_{2}}\zeta_{\vec{k}_{3}}\right]\Big\rangle&\sim& i\,\sum_{\vec{k}_{1}}\text{Im}\Big[P_{\zeta}^{-1}(k_{1})\langle Q_{\omega}\,\zeta_{\vec{k}_{1}}\rangle \langle \zeta_{\vec{k}_{1}}\zeta_{\vec{k}_{2}}\zeta_{\vec{k}_{3}}\rangle+P_{\gamma}^{-1}(k_{1})\langle Q_{\omega}\,\gamma_{ij,\vec{k}_{1}}\rangle \langle \gamma_{ij,\vec{k}_{1}}\zeta_{\vec{k}_{2}}\zeta_{\vec{k}_{3}}\rangle\nonumber\\&+&P_{\tilde\sigma}^{-1}(k_{1})\langle Q_{\omega}\,\tilde\sigma_{\vec{k}_{1}}\rangle \langle \tilde\sigma_{\vec{k}_{1}}\zeta_{\vec{k}_{2}}\zeta_{\vec{k}_{3}}\rangle\Big]\,.
\end{eqnarray}
The l.h.s. of the equation just above may be obtained by simply writing explicitly the linear part of the transformation for $\zeta$. This then gives precisely the well-known leading order contribution \cite{Maldacena:2002vr} to the tensor-scalar-scalar correlator (first term on the r.h.s. of Eq.~(\ref{uuu}))
\begin{equation}
 \big\langle \left[Q_{\omega},\zeta_{\vec{k}_{2}}\zeta_{\vec{k}_{3}}\right]\big\rangle\sim i\,    S_{ij}\sum_{a=2}^{3}k_{a}^{i}\frac{\partial}{\partial k_{a}^{j}}\big\langle\zeta_{\vec{k}_{2}}\zeta_{\vec{k}_{3}}\big\rangle^{}_{}\,.
\end{equation}
Moving on to the r.h.s. of (\ref{reee}), the first contribution, proportional to $\langle Q_{\omega}\zeta_{\vec{k}}\rangle$, vanishes because $\zeta$ transforms linearly under the tensor symmetry. Using Eq.~(\ref{rrtt}), the second contribution gives
\begin{equation}
i\,\sum_{\vec{k}_{1}}\text{Im}\left[P_{\gamma}^{-1}(k_{1})\Big\langle Q_{\omega}\,\gamma_{ij,\vec{k}_{1}}\Big\rangle \Big\langle \gamma_{ij,\vec{k}_{1}}\zeta_{\vec{k}_{2}}\zeta_{\vec{k}_{3}}\Big\rangle\right]\sim  i\lim_{\vec{k}_{1}\rightarrow 0}P_{\gamma}^{-1}(k_{1})S_{ij}\Big\langle \gamma_{ij,\vec{k}_{1}}\zeta_{\vec{k}_{2}}\zeta_{\vec{k}_{3}}\Big\rangle\,,
\end{equation}
which corresponds to the l.h.s. of Eq.~(\ref{uuu}).
A violation of the consistency conditions can therefore only arise from the third contribution to Eq.~(\ref{reee}). This is given by
\begin{equation}\label{rrrrr}
\langle Q_{\omega}\,\tilde\sigma_{\vec{k}_{1}}\rangle \langle \tilde\sigma_{\vec{k}_{1}}\zeta_{\vec{k}_{2}}\zeta_{\vec{k}_{3}}\rangle\sim       \left( \langle Q_{\omega}\sigma \rangle -\frac{\langle Q_{\omega}\zeta \rangle   \langle \zeta\,\sigma \rangle}{P_{\zeta}}  -\frac{\langle Q_{\omega}\gamma \rangle \langle \gamma\,\sigma  \rangle}{P_{\gamma}}  \right)  \langle \sigma\,\zeta^{2} \rangle +\cdot\cdot\cdot \,.
\end{equation}
Notice that the dots in Eq.(\ref{rrrrr}) stem from our expanding the state $|\tilde\sigma \rangle$ as a function of the more ``physical" $|\sigma \rangle$; the expansion is regulated by the small parameter $\epsilon(\nu)^2= |\langle \zeta \sigma\rangle|^2/(P_{\zeta}\,P_{\sigma})$. The first two terms on the r.h.s. of (\ref{rrrrr}) are again equal to zero from the linear action of the tensor symmetry on $\zeta$ and $\sigma$. The third contribution is zero in the case of QsF inflation because rotational invariance requires $\langle \sigma \gamma \rangle = 0$ . This goes to show that a ccs-violating contribution to the \textsl{tss} correlation can  be produced from a non-zero tensor-isocurvature mode correlation. While it is apparent that Lorentz invariance forbids any direct tensor-isocurvature quadratic coupling to arise from the Lagrangian in (\ref{21}), the validity of our argument also applies  {at loop level: the absence of any source breaking rotational invariance in the model ensures that the tensor-isocurvature correlation is equal to zero. Symmetry considerations indeed require the following
\begin{equation}
\langle  \gamma_{ij,\vec{k}}\,\delta\sigma_{\vec{p}}\rangle=\delta^{(3)}(\vec{k}+\vec{p})\mathcal{F}(k)\mathcal{Z}_{ij}\,,
\end{equation}
where $\mathcal{Z}_{ij}$ is a tensor, independent from the metric and the wave vector $\vec{k}$, enjoying the same properties as $\gamma_{ij}$ (symmetric, traceless, transverse). A consequence of this result for QsF is that also diagrams as the one in Fig.~(\ref{fig33}), which one would naively expect to be only mildly suppressed w.r.t. tree-level contributions\footnote{{Before relying on symmetry arguments, let us consider the case when $\sigma$ self-couplings are not much smaller than unity, that is $V^{'''}/H\lesssim \mathcal{O}(1)$. One may expect diagrams with isocurvature fluctuations running in the loop not to be significantly suppressed in comparison with tree-level contributions such as those in Fig.~\ref{fig_new_2}(h,i). The diagram in Fig.~\ref{fig33}, for instance, would naively have a squeezed limit amplitude
\begin{equation}\label{loop}
\Big\langle \gamma^{p}_{\vec{k}_{1}}\,\zeta_{\vec{k}_{2}}\,\zeta_{\vec{k}_{3}}\Big\rangle_{loop}\sim \left(\frac{V^{'''}}{H}\right)^{2} \left(\frac{\dot{\theta}}{H}\right)^{2}P_{\gamma}(k_{1})P_{\zeta}(k_{2})\left(\frac{k_{1}}{k_{2}}\right)^{3/2-\nu}\,.
\end{equation}
However, for such a diagram to be non-zero one would need to introduce some degree of anisotropy in the model.}}}, are null for symmetry reasons.\\\indent To conclude, a detection of a \textsl{tss} correlation in the limit of a soft tensor mode may point towards massive isocurvature fields during inflation, but not in the minimal set-up analysed in this work, where the massive isocurvature fields are scalars and the background is isotropic. One may break statistical isotropy by introducing vector degrees of freedom. It would also be interesting to explore scenarios with additional tensor modes. It is worth investigating possible extensions, along these lines, of the minimal set-up of QsF inflation, and assess whether the ensemble of constraints on the theory would allow for an observable fossil signal.\\ }

Within the minimal QsF model, it is also interesting to ask whether other correlators involving tensor modes may produce a violation of ccs. Consider for instance the scalar-tensor-tensor (\textsl{stt}) bispectrum. In the limit of a soft curvature fluctuation, one can follow the arguments outlined in this section for the \textsl{tss} correlation. These show that a necessary conditions for the \textsl{stt} correlation to violate ccs is that of $\zeta$ being sourced by a long-wavelength isocurvature mode. This is the case for the diagrams represented in Fig.~\ref{fig20}. A typical contribution to the latter will display a $\left(k_L / k_S \right)^{\# \nu}$ dependence in the squeezed limit which is not captured by the standard ccs.

The symmetry of interest for the squeezed \textsl{stt} correlator is a dilatation $x^{i}\rightarrow (1+\lambda)\, x^{i} $, under which the curvature fluctuation transforms as $\zeta\rightarrow \zeta+\lambda\,(1+\vec{x}\cdot\partial_{\vec{x}}\zeta)$. From the transformation of tensor modes under dilatation, $\delta_{d}\,\gamma_{ij}\sim\lambda \,x^{m}\partial_{m} \gamma_{ij}$ , one finds:
\begin{equation}\label{lhhh}
\big\langle \left[Q_{d},\gamma_{\vec{k}_{2}}^{s_{2}}\gamma_{\vec{k}_{3}}^{s_{3}}\right]\big\rangle=-i\Big\langle \delta_{d}\left(\gamma_{\vec{k}_{2}}^{s_{2}}\gamma_{\vec{k}_{3}}^{s_{3}}\right)\Big\rangle\sim i\, \lambda\left(3+\sum_{a=2,3}\vec{k}_{a}\cdot\frac{\partial}{\partial\vec{k}_{a}}\right)\Big\langle \gamma_{\vec{k}_{2}}^{s_{2}}\gamma_{\vec{k}_{3}}^{s_{3}}\Big\rangle\,,
\end{equation}
{which corresponds to the usual expression for the squeezed \textsl{stt} correlator \cite{Maldacena:2002vr}.} Similarly to Eq.~(\ref{1pa}), one can expand the l.h.s. of Eq.~(\ref{lhhh}) and introduce a complete set of states
\begin{equation}
\big\langle \left[Q_{d},\gamma_{\vec{k}_{2}}\gamma_{\vec{k}_{3}}\right]\big\rangle=\sum_{n,\vec{k}}\big\langle Q_{d}| n_{\vec{k}}\big\rangle\big\langle n_{\vec{k}} | \gamma_{\vec{k}_{2}}\gamma_{\vec{k}_{3}}\big\rangle-\sum_{n,\vec{k}}\big\langle \gamma_{\vec{k}_{2}}\gamma_{\vec{k}_{3}} | n_{\vec{k}}\big\rangle\big\langle n_{\vec{k}} | Q_{d}\big\rangle\,.
\end{equation}
As before, we consider one-particle states for $\zeta$, $\gamma$ and $\tilde\sigma$. One finds 
\begin{equation}
\Big\langle \left[Q_{q},\gamma^{2}\right] \Big\rangle \sim i\,\text{Im}\left[P_{\zeta}^{-1}\langle  Q_{d}\,\zeta  \rangle   \langle  \zeta\,\gamma^2  \rangle+P_{\gamma}^{-1}\langle Q_{d}\,\gamma   \rangle   \langle  \gamma^{3}  \rangle+P_{\tilde\sigma}^{-1}\langle Q_{d}\,\tilde\sigma   \rangle    \langle  \tilde\sigma\gamma^2  \rangle\right]\,.
\end{equation}
The $\langle  Q_{d}\,\zeta \rangle$ term gives a contribution proportional to $\langle \zeta_{\vec{q}_{}} \,\gamma_{\vec{k}_{2}} \gamma_{\vec{k}_{3}} \rangle$. The $\langle  Q_{d}\,\gamma \rangle$ term is zero from tensor modes transforming linearly under dilatation. A non-zero contribution originates from the $\langle  Q_{d}\,\tilde\sigma \rangle$ term, in the presence of a correlation between isocurvature and curvature modes, which is the case for QsF inflation. Schematically this term gives
\begin{eqnarray}
\label{3.29}
\langle  Q_{d}\,\tilde\sigma \rangle \langle  \tilde\sigma\gamma^2\rangle &\sim& \left(\langle  Q_{d}\,\sigma \rangle-P_{\zeta}^{-1}\langle  Q_{d}\,\zeta \rangle \langle  \zeta\,\sigma \rangle-P_{\gamma}^{-1}\langle  Q_{d}\,\gamma \rangle \langle  \gamma\,\sigma \rangle\right)\langle \tilde\sigma\gamma^2\rangle\nonumber\\&\sim& \lambda\,i\,\delta(\vec{q})P_{\zeta}^{-1}\langle \zeta \sigma \rangle\langle \sigma\gamma^2\rangle+\cdot\cdot\cdot \; ,
\label{hello}
\end{eqnarray}
where the first term in the first line on the r.h.s. of Eq.~(\ref{hello}) is again zero because $\sigma$ itself, as a scalar and in general a non-gravitational degree of freedom, transforms linearly under the symmetry.
The final result is that the consistency-condition violating contributions for the squeezed \textsl{stt} originates from:
\begin{equation}
P_{\sigma}^{-1}(k_{1})\Big\langle \zeta_{\vec{k}_{1}}\sigma_{\vec{k}_{1}}  \Big\rangle^{'}\Big\langle \sigma_{\vec{k}_{1}} \gamma_{\vec{k}_{2}}^{s_{2}}\gamma_{\vec{k}_{3}}^{s_{3}}   \Big\rangle^{'} \subset   \Big\langle \zeta_{\vec{k}_{1}}\gamma_{\vec{k}_{2}}^{s_{2}}\gamma_{\vec{k}_{2}}^{s_{3}}\Big\rangle^{'} \,, 
\end{equation}
and in Fig. 6 we represent some of such diagrams.

\begin{figure}
\begin{center}
\includegraphics[scale=0.35]{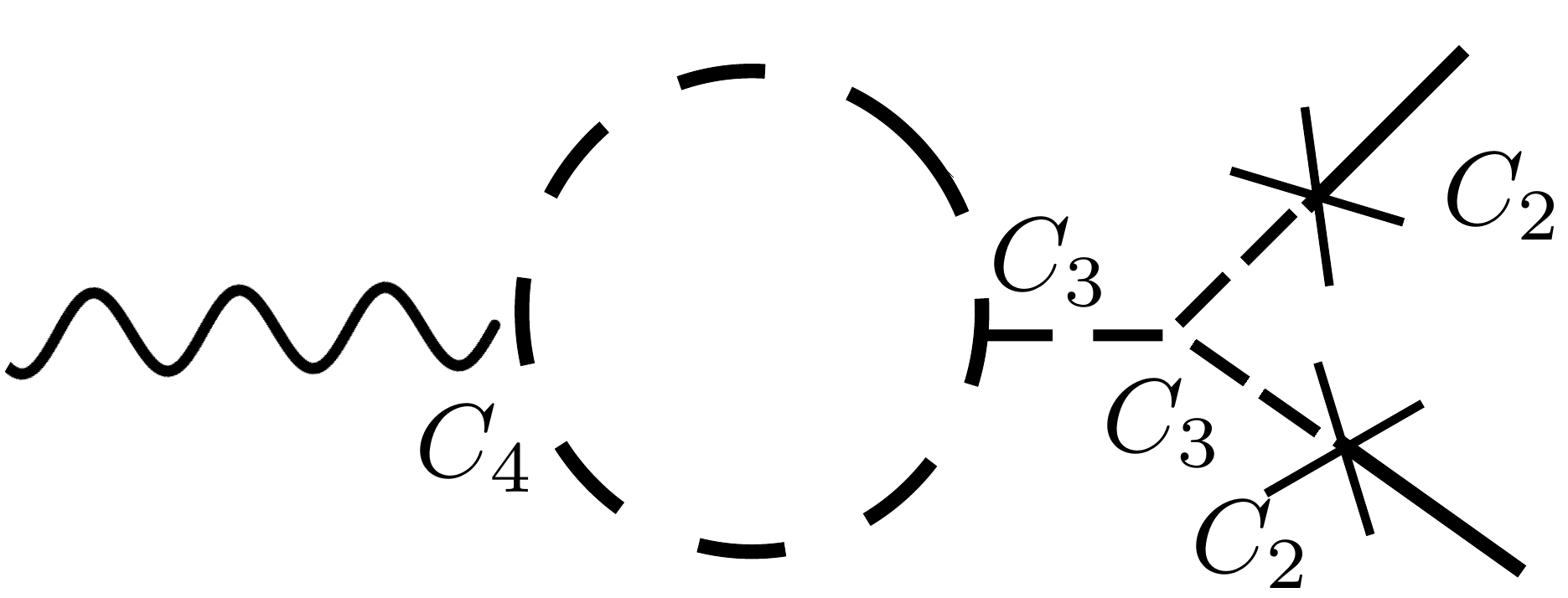}
\caption{Example of loop diagram contribution to the \textsl{tss} bispectrum. In QsF inflation any diagram that invoves a correlation between one tensor mode and one isocurvature mode is equal to zero for symmetry reasons.}
\label{fig33}
\end{center}
\end{figure}

\begin{figure}
\begin{center}
\includegraphics[scale=0.35]{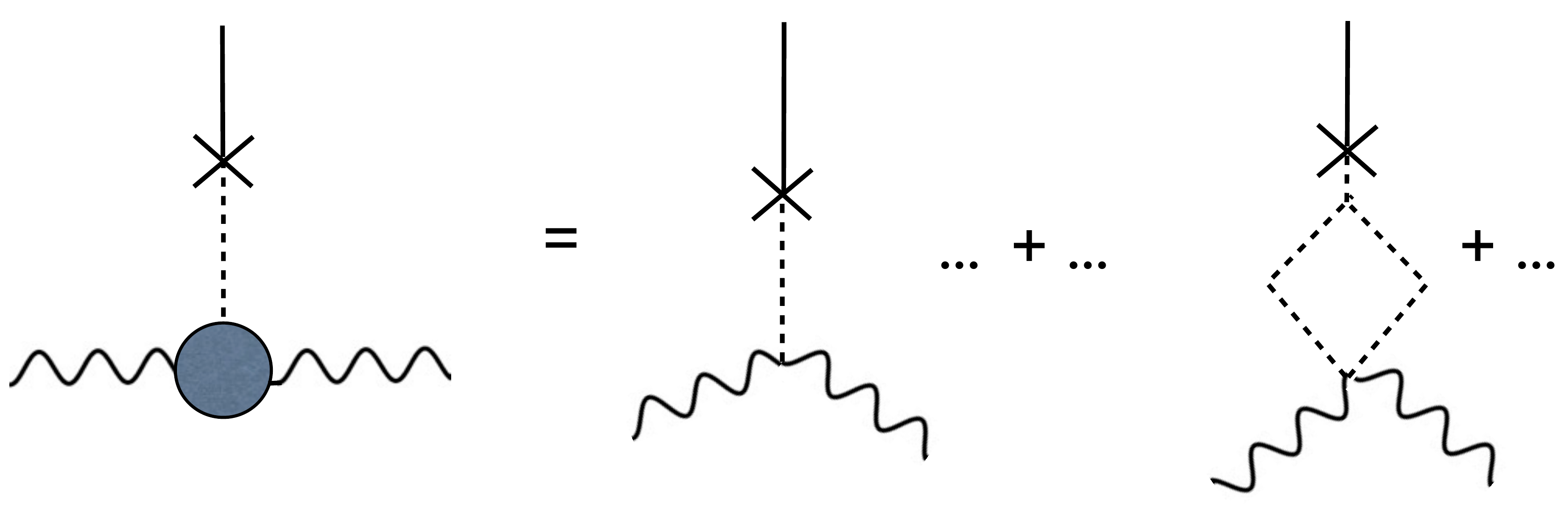}
\caption{Representation of ccs-violating contributions to the $stt$ correlation. The soft scalar curvature mode is directly sourced by a (soft) isocurvature mode. }
\label{fig20}
\end{center}
\end{figure}

\noindent It would be interesting to identify possible signatures of \textsl{stt} correlation functions in QsF inflation and in more general models with fields with $m\simeq \mathcal{O}(H)$. We leave this for future work.

\section{Conclusions}
\label{conclusions}
The combination of CMB and LSS data at our current (or
near-future) disposal represents an unprecedented source of
information on the physics of the early Universe. Most of the
cosmological evolution probed takes place at very high energies,
much higher than those one can hope to reach through particle
colliders.  Cosmological probes then might well be our best
chance at an observational window on UV phenomena.  Naturally,
unravelling the cosmological information and, most notably,
mapping it to operators in the Lagrangian of the underlying
theory, is a complicated endeavour.

Many successful efforts in this direction already populate the
literature, and Ref.~\cite{Arkani-Hamed:2015bza} recently
systematically classified the properties of inflationary
curvature correlation functions in the presence of additional fields with
Hubble-scale masses (see also \cite{Kehagias:2015jha} for a very general analysis on inflationary correlators in the presence of massive fields). The oscillatory (or, alternatively, power-law)
behavior of correlators in the squeezed limit provides a measure
of the mass of the fields while the angular dependence encodes
knowledge about their spin \cite{Jeong:2012df}.  Here as
elsewhere the study of the squeezed limit of correlations
functions proves crucial for investigating new degrees of
freedom during inflation. One may well see part of the present work as complementing the one in \cite{Arkani-Hamed:2015bza} in that we study the effect of massive fields on correlators that also contain gravitons.

Here we investigated the observability of squeezed three-point correlation functions involving both curvature and tensor fluctuations in QsF inflation. The latter is a phenomenological inflationary scenario comprising one or more fields with mass of order Hubble. In order to assess the observability of such signals, the first step is to verify whether these bispectra violate single-clock consistency conditions. We checked from general principles that, in QsF inflation, this is the case for correlations among one (soft) curvature mode and two tensor modes (\textsl{stt}). The latter opens up the possibility of new late-time observable effects in CMB fluctuations or in the galaxy distribution. It would be interesting to further investigate such a possibility in this as well as other inflationary models that include $m\sim H$ fields. \\
\indent An observable of great interest, as pointed out in previous works in the context of other inflationary models \cite{Dimastrogiovanni:2014ina}, is the \textsl{tss} correlation (one long-wavelength tensor and two scalar modes). In \cite{Jeong:2012df}-\cite{Dai:2013kra} it was shown how, in the presence of such a primordial correlation, a soft super-Hubble tensor mode would induce a power quadrupole, whereas a sub-Hubble tensor mode would introduce local departures from statistical isotropy in the form of an off-diagonal correlation between different Fourier modes of the density field. The latter could be used as a probe for the amplitude of tensor modes and, for a given value of the Hubble parameter during inflation, one can estimate the size of the survey necessary to access the physics at that scale H. This is not the most direct way of probing the existence of tensor modes. Rather, it is a specific probe that relies on, and therefore crucially contains information about,  the existence of a non-Gaussian signature arising from the multi-field and non-single clock nature of the QsF model. It is in other words a route to seeking critical information on the nature of the inflationary dynamics. \\ 
\indent In this paper, we show that violation of consistency conditions for the \textsl{tss} correlation would occur in the presence of a non-zero two-point correlation between the tensor mode and the massive isocurvature. Symmetry requirements (statistical isotropy) in the minimal QsF inflation realization considered in this paper forbid such a coupling. However, tensor-isocurvature correlations may well be generated within non-minimal extensions of QsF. This may be the case if one breaks rotational invariance e.g. by extending the particle content to include additional vector or tensor degrees of freedom. These are interesting possibilities, worth exploring.

 One may also wonder how these results may be extended to the fully
multi-field case (in the sense of non-decaying light degrees of freedom alongside the inflaton field). As mentioned, a high degree of model
dependence is intrinsic to these setups. Nevertheless, the main
properties of QsF inflation we have employed here, also characterizes the dynamics of multiple light fields. We leave this to future work.\\

\section*{Acknowledgments}

It is a pleasure to thank Xingang Chen and Mohammad Hossein Namjoo for
collaboration at the early stages of this project as well as for
fruitful discussions. We are also indebted to Paolo Creminelli, Donghui Jeong and
Jared Kaplan for illuminating conversations. This work was
supported at ASU by the Department of Energy, at JHU by NSF
Grant No. 0244990, NASA NNX15AB18G, the John Templeton
Foundation, and the Simons Foundation, and at Stanford by NSF
grant PHY-1068380. ED and MF are delighted to thank the
Cosmology group at JHU for very warm hospitality whilst parts of
this work were being completed.

\appendix

\section{Computation of $\delta\sigma$-mediated \textsl{tss} correlation in the squeezed limit}
\label{diagramf}

The expectation value of an operator $\Theta$ at time $t$ can be
computed using the in-in formula,
\ba
\left\langle \Theta(t) \right\rangle=\left\langle 0\,\Big| \left[\bar{T}\,\exp\left(i\int_{t_{0}}^{t}dt^{'}H_{I}(t^{'})\right)\right] \Theta_{I}(t)\left[T\,\exp\left(-i\int_{t_{0}}^{t}dt^{''}H_{I}(t^{''})\right)\right] \Big|\,0 \right\rangle\,,
\ea
where $H_{I}$ is the interaction Hamiltonian. For a three-vertex
diagram one finds 
\ba\label{32}
\left\langle \Theta(t) \right\rangle&=&-2\,\mathcal{I}\left[\int_{t_{0}}^{t}dt_{1}\int_{t_{0}}^{t_{1}}dt_{2}\int_{t_{0}}^{t_{2}}dt_{3}\left\langle  \Theta_{I}(t)H_{I}(t_{1})H_{I}(t_{2})H_{I}(t_{3})\right\rangle\right]\\\label{33}
&+&2\,\mathcal{I}\left[\int_{t_{0}}^{t}d\tilde{t}_{1}\int_{t_{0}}^{t}dt_{1}\int_{t_{0}}^{t_{1}}dt_{2}\left\langle H_{I}(\tilde{t}_{1}) \Theta_{I}(t)H_{I}(t_{1})H_{I}(t_{2})\right\rangle\right]\,,
\ea
where $\mathcal{I}$ stands for the imaginary part.

For an operator
$\Theta\sim\gamma_{\vec{k}_{1}}\delta\theta_{\vec{k}_{2}}\delta\theta_{\vec{k}_{3}}$
and an interaction Hamiltonian $H_{I}\equiv H_{2}+H_{3}$, where
$H_{2}\sim \delta\sigma\delta\theta^{'}$ and $H_{3}\sim
\gamma_{ij}\partial_{i}\delta\sigma\partial_{j}\delta\sigma$,
from the right-hand side of Eq.~(\ref{32}), three terms arise (the same
goes for Eq.~(\ref{33})). The number of permutations for each of the
six different terms is $4$, so there is a total of $24$
terms. The right hand side of Eq.~(\ref{32}) is then given (modulo
an overall factor) by the sum of the following terms:
\ba\label{3.6}
&&-2 \epsilon_{ij}^{\lambda}k_{2i}k_{3j}\,\gamma_{k_{1}}(0)u_{k_{2}}(0)u_{k_{3}}(0)\times\mathcal{I}\Bigg[\int_{-\infty}^{0}d\tau_{1}\,
a^{2}(\tau_{1})\,\gamma_{k_{1}}^{*}(\tau_{1})v_{k_{2}}(\tau_{1})v_{k_{3}}(\tau_{1})\\&&\quad\quad\times
\int_{-\infty}^{\tau_{1}}d\tau_{2}\,a^{3}(\tau_{2})\,v_{k_{2}}^{*}(\tau_{2})u_{k_{2}}^{'*}(\tau_{2})\int_{-\infty}^{\tau_{2}}d\tau_{3}\,a^{3}(\tau_{3})\,v_{k_{3}}^{*}(\tau_{3})u_{k_{3}}^{'*}(\tau_{3})\Bigg] + 3\,\,{\rm perms}.\nonumber
\ea
\ba
\label{3.7}
&&-2 \epsilon_{ij}^{\lambda}k_{2i}k_{3j}\,\gamma_{k_{1}}(0)u_{k_{2}}(0)u_{k_{3}}(0)\times\mathcal{I}\Bigg[\int_{-\infty}^{0}d\tau_{1}\,a^{3}(\tau_{1})\,v_{k_{2}}^{}(\tau_{1})u_{k_{2}}^{'*}(\tau_{1})\\&&\quad\quad\times\int_{-\infty}^{\tau_{1}}d\tau_{2}\,  a^{2}(\tau_{2})\,\gamma_{k_{1}}^{*}(\tau_{2})v_{k_{2}}^{*}(\tau_{2})v_{k_{3}}^{}(\tau_{2}) \int_{-\infty}^{\tau_{2}}d\tau_{3}\, a^{3}(\tau_{3})\,v_{k_{3}}^{*}(\tau_{3})u^{'*}_{k_{3}}(\tau_{3})\Bigg] + 3\,\,{\rm perms}.\,\,\nonumber
\ea
\ba
\label{3.8}
&&-2 \epsilon_{ij}^{\lambda}k_{2i}k_{3j}\,\gamma_{k_{1}}(0)u_{k_{2}}(0)u_{k_{3}}(0)\times\mathcal{I}\Bigg[\int_{-\infty}^{0}d\tau_{1}\,a^{3}(\tau_{1})\,v_{k_{2}}^{}(\tau_{1})u^{'*}_{k_{2}}(\tau_{1})\\&&\quad\quad\times\int_{-\infty}^{\tau_{1}}d\tau_{2}\, a^{3}(\tau_{2})\, v_{k_{3}}^{}(\tau_{2})u_{k_{3}}^{'*}(\tau_{2}) \int_{-\infty}^{\tau_{2}}d\tau_{3}\,a^{2}(\tau_{3})\,\gamma_{k_{1}}^{*}(\tau_{3})v_{k_{2}}^{*}(\tau_{3})v_{k_{3}}^{*}(\tau_{3})\Bigg] + 3\,\,{\rm perms}.\nonumber
\ea

Eq.~(\ref{33}) is given by the sum of the following terms
\ba\label{ppp}
&&2 \epsilon_{ij}^{\lambda}k_{2i}k_{3j}\,\gamma_{k_{1}}^{*}(0)u_{k_{2}}^{}(0)u_{k_{3}}^{}(0)\times\mathcal{I}\Bigg[\int_{-\infty}^{0}d\tilde{\tau}_{1}\, a^{2}(\tilde{\tau}_{1})\,\gamma_{k_{1}}^{}(\tilde{\tau}_{1})v_{k_{2}}^{}(\tilde{\tau}_{1})v_{k_{3}}^{}(\tilde{\tau}_{1}) \\&&\quad\quad\times\int_{-\infty}^{0}d\tau_{1}\,a^{3}(\tau_{1})\,v_{k_{2}}^{*}(\tau_{1})u_{k_{2}}^{'*}(\tau_{1})\int_{-\infty}^{\tau_{1}}d\tau_{2}\,a^{3}(\tau_{2})\,v_{k_{3}}^{*}(\tau_{2})u_{k_{3}}^{'*}(\tau_{2})\Bigg] + 3\,\,{\rm perms}.\nonumber
\ea
\ba
\label{ppp1}
&&2 \epsilon_{ij}^{\lambda}k_{2i}k_{3j}\,\gamma_{k_{1}}^{}(0)u_{k_{2}}^{*}(0)u_{k_{3}}^{}(0)\times\mathcal{I}\Bigg[\int_{-\infty}^{0}d\tilde{\tau}_{1}\, a^{3}(\tilde{\tau}_{1})\,v_{k_{2}}^{}(\tilde{\tau}_{1})u_{k_{2}}^{'}(\tilde{\tau}_{1}) \\&&\quad\quad\times\int_{-\infty}^{0}d\tau_{1}\,a^{2}(\tau_{1})\,\gamma_{k_{1}}^{*}(\tau_{1})v_{k_{2}}^{*}(\tau_{1})v_{k_{3}}^{}(\tau_{1})\int_{-\infty}^{\tau_{1}}d\tau_{2}\,a^{3}(\tau_{2})\,v_{k_{3}}^{*}(\tau_{2})u_{k_{3}}^{'*}(\tau_{2})\Bigg] + 3\,\,{\rm perms}.\nonumber
\ea
\ba
\label{ppp2}
&&2 \epsilon_{ij}^{\lambda}k_{2i}k_{3j}\,\gamma_{k_{1}}^{}(0)u_{k_{2}}^{*}(0)u_{k_{3}}^{}(0)\times\mathcal{I}\Bigg[\int_{-\infty}^{0}d\tilde{\tau}_{1}\,a^{3}(\tilde{\tau}_{1})\,v_{k_{2}}^{}(\tilde{\tau}_{1})u^{'}_{k_{2}}(\tilde{\tau}_{1})\\\label{3.11}&&\quad\quad\times\int_{-\infty}^{0}d\tau_{1}\,a^{3}(\tau_{1})\,v_{k_{3}}^{}(\tau_{1})u_{k_{3}}^{'*}(\tau_{1})\int_{-\infty}^{\tau_{1}}d\tau_{2}\,a^{2}(\tau_{2})\,\gamma_{k_{1}}^{*}(\tau_{2})v_{k_{2}}^{*}(\tau_{2})v_{k_{3}}^{*}(\tau_{2})\Bigg] + 3\,\,{\rm perms}.\nonumber
\ea

In the equations above one uses the Fourier-mode decomposition,
\ba
\delta\theta(\tau,\vec{x})&=&\int \frac{d^{3}k}{(2\pi)^{3}}\, e^{-i\vec{k}\cdot\vec{x}} \left[a_{\vec{k}}\,u_{k}(\tau)+a^{\dagger}_{-\vec{k}}\,u^{*}_{k}(\tau)\right],\\
\delta\sigma(\tau,\vec{x}) &=&   \int \frac{d^{3}k}{(2\pi)^{3}}\, e^{-i\vec{k}\cdot\vec{x}} \left[b_{\vec{k}}\,v_{k}(\tau)+b^{\dagger}_{-\vec{k}}\,v^{*}_{k}(\tau)\right],\\\label{14}
\gamma_{ij}(\tau,\vec{x})&=&\int \frac{d^{3}k}{(2\pi)^{3}}\, e^{-i\vec{k}\cdot\vec{x}}\sum_{\lambda=\pm}\epsilon_{ij}^{\lambda}(\hat{k}) \left[c^{\lambda}_{\vec{k}}\,\gamma_{k}(\tau)+\left(c^{\lambda}_{-\vec{k}}\right)^{\dagger}\,\gamma^{*}_{k}(\tau)\right],
\ea
for the fields. The mode functions are given by
\ba
u_{k}(\tau)=\frac{H}{R\sqrt{2k^{3}}}\left(1+i k \tau\right)e^{-i k \tau},\quad\quad
\gamma_{k}(\tau)=\frac{H}{M_{P}\sqrt{k^{3}}}\left(1+i k \tau\right)e^{-i k \tau},\\
v_{k}(\tau)=-i e^{i\left(\nu+\frac{1}{2}\right)\frac{\pi}{2}}\frac{\sqrt{\pi}}{2} H \left(-\tau\right)^{3/2}H_{\nu}^{(1)}(-k \tau)\,,\quad\quad\quad\,\,\,
\ea
where $\nu\equiv\sqrt{9/4-(m/H)^{2}}$ and $m^{2}/H^{2}\leq
9/4$. Consider the first permutation of Eq.~(\ref{3.6}) and
replace the expressions for the mode functions.  One then finds,
\ba\label{int}
-\epsilon_{ij}^{\lambda}k_{2i}k_{3j}\frac{\pi^{2}}{2^{5}}\frac{H^{2}}{M_{P}^{2}R^{4}}\frac{1}{k_{1}^{3}k_{2}k_{3}}\times \mathcal{I}\Bigg[ \int_{-\infty}^{0}d\tau_{1} \int_{-\infty}^{\tau_{1}}d\tau_{2} \int_{-\infty}^{\tau_{2}}d\tau_{3}  \frac{(-\tau_{1})}{(-\tau_{2})^{1/2}(-\tau_{3})^{1/2}} \quad\quad\quad\quad\quad\\ \left(1-i k_{1}\tau_{1}\right)e^{i k_{1}\tau_{1}}  H_{\nu}^{(1)}(-k_{2}\tau_{1})  H_{\nu}^{(1)}(-k_{3}\tau_{1}) e^{i k_{2}\tau_{2}} H_{\nu}^{(2)}(-k_{2}\tau_{2}) e^{i k_{3}\tau_{3}} H_{\nu}^{(2)}(-k_{3}\tau_{3}) \Bigg]. \nonumber
\ea
Working in the squeezed limit $k_{1}\ll k_{2}, \, k_{3}$, it is
convenient to perform a change of variables introducing
$x_{i}\equiv k_{3}\tau_{i}$. Then the result in Eq.~(\ref{int})
becomes
\ba\label{int22}
-\epsilon_{ij}^{\lambda}k_{2i}k_{3j}\frac{\pi^{2}}{2^{5}}\frac{H^{2}}{M_{P}^{2}R^{4}}\frac{1}{k_{1}^{3}k_{2}k_{3}^{4}}\times \mathcal{I}\Bigg\{\int_{-\infty}^{0}dx_{1}   \int_{-\infty}^{x_{1}}dx_{2}  \int_{-\infty}^{x_{2}}dx_{3}\frac{(-x_{1})}{(-x_{2})^{1/2}(-x_{3})^{1/2}}\quad\quad\quad\quad\quad\quad\quad\quad\\\left[1-i \left(\frac{k_{1}}{k_{3}}\right)x_{1}\right] e^{i\left(\frac{k_{1}}{k_{3}}\right)x_{1}}  H_{\nu}^{(1)}\left[-\left(\frac{k_{2}}{k_{3}}\right)x_{1}\right]H_{\nu}^{(1)}[-x_{1}] e^{i \left(\frac{k_{2}}{k_{3}}\right)x_{2}} H_{\nu}^{(2)}\left[-\left(\frac{k_{2}}{k_{3}}\right)x_{2}\right] e^{i x_{3}} H_{\nu}^{(2)}\left[-x_{3}\right] \Bigg\}\nonumber
\ea
Next, one can use the approximation $k_{2}\simeq k_{3}$ (valid
in the squeezed limit) in the first and in the third Hankel
functions of the previous expression as well as in the second
exponential. The following approximation is also allowed 
\ba\label{pr}
\left[1-i \left(\frac{k_{1}}{k_{3}}\right)x_{1}\right] e^{i\left(\frac{k_{1}}{k_{3}}\right)x_{1}}\simeq 1.
\ea
To see why notice that $k_{1}\ll k_{3}$.  Therefore in order to
have $(k_{1}/k_{3}) |x_{1}|\geq 1$ one would need $|x_{1}|\gg
1$, but since $x_{1}$ is the argument of Hankel functions in
Eq.~(\ref{int22}), then in the limit of very large $|x_{1}|$ these
functions would be rapidly oscillating and suppress the value of
the integrals.

With these approximations, the final result for
Eq.~(\ref{int22}) then becomes
\ba\label{pr0}
-\epsilon_{ij}^{\lambda}\hat{k}_{2i}\hat{k}_{3j}\frac{\pi^{2}}{2^{5}}\frac{H^{2}}{M_{P}^{2}R^{4}}\frac{1}{k_{1}^{3}k_{3}^{3}}\times \mathcal{I}\Bigg\{\int_{-\infty}^{0}dx_{1}   \int_{-\infty}^{x_{1}}dx_{2}  \int_{-\infty}^{x_{2}}dx_{3}\frac{(-x_{1})}{(-x_{2})^{1/2}(-x_{3})^{1/2}}\quad\quad\quad\quad\quad\\ \left(H_{\nu}^{(1)}[-x_{1}] \right)^{2}e^{i x_{2}} H_{\nu}^{(2)}\left[-x_{2}\right] e^{i x_{3}} H_{\nu}^{(2)}\left[-x_{3}\right] \Bigg\},\nonumber
\ea
where unit vectors $\hat{k}\equiv \vec{k}/k$ have been
introduced.  With a similar procedure, the first permutation in
Eq.~(\ref{ppp}) gives
\ba\label{pr1}
\epsilon_{ij}^{\lambda}\hat{k}_{2i}\hat{k}_{3j}\frac{\pi^{2}}{2^{5}}\frac{H^{2}}{M_{P}^{2}R^{4}}\frac{1}{k_{1}^{3}k_{3}^{3}}\times \mathcal{I}\Bigg\{\int_{-\infty}^{0}d\tilde{x}_{1}   \int_{-\infty}^{0}dx_{1}  \int_{-\infty}^{x_{1}}dx_{2}\frac{(-\tilde{x}_{1})}{(-x_{1})^{1/2}(-x_{2})^{1/2}}\quad\quad\quad\quad\quad\\ \left(H_{\nu}^{(1)}[-\tilde{x}_{1}] \right)^{2}e^{i x_{1}} H_{\nu}^{(2)}\left[-x_{1}\right] e^{i x_{2}} H_{\nu}^{(2)}\left[-x_{2}\right] \Bigg\}.\nonumber
\ea
The total $\langle\gamma\delta\theta\delta\theta\rangle$ is
obtained by adding up all the permutations listed in
Eqs.~(\ref{3.6})--(\ref{ppp2}). Unlike what happens for the
$\langle\delta\theta\delta\theta\delta\theta\rangle$ correlation
in the squeezed limit \cite{Chen:2009zp}, for the \textsl{tss}
correlator one expects that all of the permutations will have
the same momentum dependence as in Eqs.~(\ref{pr0}) and
(\ref{pr1}); e.g. $\sim k_{L}^{-3}k_{S}^{-3}$ ($k_{L}$ and
$k_{S}$ being respectively long and short-wavelength modes).

Summing up all permutations in Eqs.~(\ref{3.6})--(\ref{3.8}),
one then finds,
\ba\label{totalalpha}
\epsilon_{ij}^{\lambda}\hat{k}_{Si}\hat{k}_{Sj}\frac{\pi^{2}}{2^{3}}\frac{H^{2}}{M_{P}^{2}R^{4}}\frac{1}{k_{L}^{3}k_{S}^{3}}\times \mathcal{I}\Bigg\{\int_{-\infty}^{0}dx_{1}   \int_{-\infty}^{x_{1}}dx_{2}  \int_{-\infty}^{x_{2}}dx_{3}\Bigg[\frac{(-x_{1})}{(-x_{2})^{1/2}(-x_{3})^{1/2}}\quad\quad\quad\quad\quad\quad\quad\quad\\ \times\left(H_{\nu}^{(1)}[-x_{1}] \right)^{2}e^{i x_{2}} H_{\nu}^{(2)}\left[-x_{2}\right] e^{i x_{3}} H_{\nu}^{(2)}\left[-x_{3}\right] +\frac{(-x_{2})}{(-x_{1})^{1/2}(-x_{3})^{1/2}} H_{\nu}^{(1)}[-x_{1}] e^{i x_{1}} H_{\nu}^{(2)}\left[-x_{2}\right]\quad\quad\nonumber\\\times H_{\nu}^{(1)}\left[-x_{2}\right] e^{i x_{3}} H_{\nu}^{(2)}\left[-x_{3}\right]+\frac{(-x_{3})}{(-x_{1})^{1/2}(-x_{2})^{1/2}}H_{\nu}^{(1)}\left[-x_{1}\right]e^{i x_{1}} H_{\nu}^{(1)}\left[-x_{2}\right] e^{i x_{2}} \left(H_{\nu}^{(2)}[-x_{3}] \right)^{2} \Bigg]\Bigg\}.\nonumber
\ea
The sum of Eqs.~(\ref{ppp}), (\ref{ppp1}) and (\ref{ppp2}) give
\ba\label{totalbeta}
& &-\epsilon_{ij}^{\lambda}\hat{k}_{Si}\hat{k}_{Sj}\frac{\pi^{2}}{2^{3}}\frac{H^{2}}{M_{P}^{2}R^{4}}\frac{1}{k_{L}^{3}k_{S}^{3}}\times \mathcal{I}\Bigg\{\int_{-\infty}^{0}d\tilde{x}_{1}   \int_{-\infty}^{0}dx_{1}  \int_{-\infty}^{x_{1}}dx_{2}\Bigg[\frac{(-\tilde{x}_{1})}{(-x_{1})^{1/2}(-x_{2})^{1/2}}\left(H_{\nu}^{(1)}[-\tilde{x}_{1}] \right)^{2}e^{i x_{1}}\nonumber\\ & &\times H_{\nu}^{(2)}\left[-x_{1}\right] e^{i x_{2}} H_{\nu}^{(2)}\left[-x_{2}\right]+\frac{(-x_{1})}{(-\tilde{x}_{1})^{1/2}(-x_{2})^{1/2}} H_{\nu}^{(1)}[-\tilde{x}_{1}] e^{-i \tilde{x}_{1}} H_{\nu}^{(2)}\left[-x_{1}\right]H_{\nu}^{(1)}\left[-x_{1}\right] e^{i x_{2}} H_{\nu}^{(2)}\left[-x_{2}\right] \nonumber \\& &+\frac{(-x_{2})}{(-\tilde{x}_{1})^{1/2}(-x_{1})^{1/2}} e^{-i \tilde{x}_{1}} H_{\nu}^{(1)}\left[-\tilde{x}_{1}\right] e^{i x_{1}} H_{\nu}^{(1)}\left[-x_{1}\right] \left(H_{\nu}^{(2)}[-x_{2}] \right)^{2}\Bigg]\Bigg\}\,.
\ea

\end{document}